# Quantum Computing in Transport Science: A Review

**Chence Niu***
Guangdong Basic Research Center of Excellence for Ecological Security and Green Development, Key Laboratory for City Cluster Environmental Safety and Green Development of the Ministry of Education, School of Ecology, Environment and Resources, Guangdong University of Technology, Guangzhou, 510006, China
Research Centre for Integrated Transport Innovation (rCITI), School of Civil and Environmental Engineering, UNSW Sydney, Sydney NSW 2052, Australia
Email: niuchence@163.com

**Elnaz Irannezhad**
Research Centre for Integrated Transport Innovation (rCITI), School of Civil and Environmental Engineering, UNSW Sydney, Sydney NSW 2052, Australia
Email: e.irannezhad@unsw.edu.au

**Casey Myers**
School of Physics, Mathematics and Computing, UWA, 35 Stirling Hwy, Crawley WA 6009
Email: casey.myers@uwa.edu.au

**Vinayak Dixit**
Research Centre for Integrated Transport Innovation (rCITI), School of Civil and Environmental Engineering, UNSW Sydney, Sydney NSW 2052, Australia
Email: v.dixit@unsw.edu.au

*Corresponding author

**Abstract**

Quantum computing, leveraging the principles of quantum mechanics, has been found to significantly enhance computational capabilities in principle, in some cases beyond classical computing limits. This paper explores quantum computing's potential to address complex, large-scale problems in transportation systems. It focuses on three principal paradigms: Gate-based quantum computing, Quantum annealing, and Quantum machine learning, which, though based on gate-based quantum computing, is treated as distinct due to its unique methods and applications. Each paradigm's foundational concepts, practical applications, and potential impacts on the field are discussed to provide a comprehensive overview of quantum computing strategies and their future implications.

1. **Introduction**

Quantum computing (QC) was first introduced in the early 1980s with the concept of quantum circuits (Benioff, 1980; Feynman, 2018). The field gained further momentum with development of Shor's algorithm (1994) and Grover's search algorithm (1996), both of which showcased QC's potential to outperform classical computing for specific tasks. Over the past three decades, QC has seen significant advancements, attributed in part to its reliance on the quantum mechanical properties of qubits, such as superposition and entanglement, which are thought to offer potential advantages over classical computing. Unlike classical bits that exist in a binary state of either 0 or 1, a quantum bit or qubit can simultaneously embody both states due to superposition. Consequently, two qubits can simultaneously represent four states: (0,0), (1,0), (0,1), and (1,1). This attribute allows for exponential scaling of state representation with additional qubits giving QC a possible significant edge over classical methods. For instance, a 20-qubit quantum computer can concurrently represent approximately 1 million ($2^{20}$) different states. A notable achievement in this field is the claim of 'Quantum Supremacy', where a 53-qubit Sycamore processor solved a problem in 200 seconds that would take a classical supercomputer approximately 10,000 years (Arute et al., 2019). This breakthrough has made QC paradigm a very attractive route for solving complex and combinatorial optimization problems, improving predictive models and enhancing decision-making processes across various domains.

Research into QC's capabilities is expanding into diverse fields, including transportation (Cooper, 2022; V. Dixit & Jian, 2022; V. V. Dixit et al., 2023; V. V. Dixit & Niu, 2023), engineering design (Chen et al., 2022), finance (Bova et al., 2021), and energy system (Ajagekar & You, 2019). In transportation, two common approaches for formulating optimisation problems in QC include:
- Hamiltonian
- Quadratic Unconstrained Binary Optimization (QUBO)[1]

Moreover, there are mainly three leading paradigms of QC algorithms currently applied in transportation research fields:
   i. Gate-based QC optimization,
   ii. Quantum Annealing (QA) optimization,
   iii. Quantum Machine Learning (QML).

QML is noted as a distinct paradigm, recognized for its unique methods and applications, despite its roots in gate-based quantum computing. Each of these paradigms brings approaches, algorithms, hardware configurations, and applications, offering different advantages depending on the problem at hand. In the following sections, we outline the foundational concepts behind these two model formulations and the three key QC paradigms, examining how they address large-scale transportation problems. Additionally, the discussion extends to various algorithmic techniques that are integral to implementing these QC solutions effectively, offering insights into their practical

---

[1] It is not entirely rigorous to separate QUBO from the Hamiltonian function; however, it is discussed here due to its significance in quantum annealing, where QUBO is transformed into a Hamiltonian that encodes the optimization problem.

integration and potential benefits.

## 2. Problem Formulation

To effectively apply quantum computing (QC) algorithms to transportation problems, classical models must be transformed into formulations that are compatible with quantum algorithms. The most common quantum formulations are Hamiltonian and Quadratic Unconstrained Binary Optimization (QUBO). This transformation is crucial because these formulations are specifically designed to be processed by quantum computers, including quantum annealers and certain gate-based quantum algorithms.

### 2.1 Hamiltonian

The Hamiltonian represents the total energy of a system and dictates how qubits evolve over time. For example, in the Quantum Approximate Optimisation Algorithm (QAOA), the optimisation problem is encoded through two key Hamiltonians: the Cost Hamiltonian and the Mixer Hamiltonian. QAOA works by alternating between applying the Cost Hamiltonian and the Mixer Hamiltonian in a sequence of quantum gates (Bourreau et al., 2022). In QA, the Hamiltonian incorporate both the objective function and constraints in a unified framework. Unlike classical optimization problems which separates the objective function from constraints, the Hamiltonian formulation integrates both elements into a single formulation. In this approach, qubits are used to represent the state of the system, allowing the Hamiltonian to encode the problem's complexity directly. For instance, the Vehicle Routing Problem (VRP), a classical NP-hard combinatorial optimisation problem, involves minimizing total travel distance while satisfying constraints such as returning to the depot. In the Hamiltonian, the VRP can be expressed as follows (Azad et al., 2023; Mohanty et al., 2023):

$$\min H_{VRP} = \min(H_A + H_B + H_C + H_D + H_E) \quad (1)$$

$$H_A = \sum_{i \to j} w_{ij} x_{ij} \quad (2)$$

$$H_B = A \sum_{i \in 1,\dots,n-1} \left(1 - \sum_{j \in source\,[i]} x_{ij}\right)^2 \quad (3)$$

$$H_C = A \sum_{i \in 1,\dots,n-1} \left(1 - \sum_{j \in target\,[i]} x_{ji}\right)^2 \quad (4)$$

$$H_D = A(k - \sum_{j \in source\,[0]} x_{0j})^2 \qquad (5)$$

$$H_E = A(k - \sum_{j \in traget\,[0]} x_{j0})^2 \qquad (6)$$

where $A > 0$ is the Lagrange multiplier. Equation (2) corresponds to the classical objective function, which aims to minimize the total travel cost. Equations (3)-(4) constitute the Node-Visit Constraint, ensuring that each node is visited exactly once. Equations (5)-(6) are the depot constraint, ensuring that all the vehicles begin from and return back to depot. Both QA and some gate-based quantum algorithms utilise Hamiltonians to define and solve problems, as will be discussed in detail in Section 3 and 4.

### 2.2 QUBO

QUBO is a widely used problem formulation for optimization problems, particularly in combinatorial optimization. In QA, the QUBO problem is transformed into an Ising Hamiltonian, where the ground state (lowest energy state) of the Hamiltonian represents the optimal solution to the QUBO problem (see Section 4.2). In QUBO, the goal is to minimize a quadratic function over binary variables, without any constraint. In the QC literature, large-scale traffic control optimization problems are frequently transformed into QUBO formulations by setting different constraint conditions and decomposing higher-order problems. For example, the QUBO formulation of airplane conflict-resolution problem is expressed as follows: (Stollenwerk et al., 2020).

$$f = \lambda \sum_{i=1}^{N_f} \left( \sum_{l=0}^{N_d} d_{i,l} - 1 \right)^2 \qquad (7)$$

where $\lambda$ is a sufficiently large penalty weight, to ensure that the state minimising the cost satisfy the condition $f = 0$. In order to encode the configuration space $\boldsymbol{d}$ as binary variables, the discrete values are employed in the form $\{\Delta_d l | l \in [0,1, \ldots, N_d]\}$, where $N_d \Delta_d$ is the maximum allowed delay. The value of $d_i$ is represented as $N_d + 1$ variables $d_{i,0}, \ldots, d_{i,N_d} \in \{0,1\}$ using a one-hot encoding:

$$d_{i,l} = \begin{cases} 1, d_i = l, \\ 0, d_i \neq l; \end{cases} \quad d_i = \Delta_d \sum_{l=0}^{N_d} l d_{i,l} \qquad (8)$$

Another example is transforming the traffic signal control problem as a QUBO formulation, as follows (Hussain et al., 2020):

$$Obj = -\lambda_1 \sum_i \sum_j C_{i,j} x_{ij}^2$$

$$-\lambda_2 \sum_i \sum_j C_{ij} x_{ij} \big[\tau_{i,a'}\lambda_3 C_{a'a} x_{a',a} + \tau_{i,b'}\lambda'_3 C_{b'b} x_{b',b} \quad (9)$$

$$+ \tau_{i,c'}\lambda_3 C_{c'c} x_{c',c} + \tau_{i,d'}\lambda'_3 C_{d'd} x_{d',d}\big] + \lambda_4 \sum_i \left[1 - \sum_j x_{i,j}\right]^2$$

where the traffic clearance cost, intersection coordination constraint and mode selection constraint are included. An artificial 6 × 6 grid map example is tested and the results show that the hybrid QA is often outperformed by purely classical algorithms like tabu search and simulated annealing. In order to address the time-dependent aspect of the problem, multiple QUBO instances are solved every 5–10 seconds, each producing a new configuration of modes across the map.

Domino et al. (2022) formulated the railway rescheduling optimization problem as a QUBO and higher-order binary optimization (HOBO) representations. Compared to QUBO, HOBO is able to deal with single, double and multi-track lines with stations and switches. The representation of HOBO problem involves minimizing a multilinear polynomial with binary variables as follows:

$$h(\mathbf{X}) = \sum_{S \subseteq V} c_S \prod_{i \in S} x_i \quad (10)$$

where $\mathbf{X}$ is the vector of all binary variables, $V = \{1, 2, \ldots, n\}$ and $c_S$ are the coefficients. The order of a HOBO is the size of the largest set $S$. Next, the HOBO is converted to QUBO representation. For example, the cubit terms $x_{i1} x_{i2} x_{i3} = 0$ is converted to $\tilde{x}_k x_{i3} = 0$, where $\tilde{x}_k = x_{i1} x_{i2}$. Then the penalty function can be defined as Equation (12) where $h(x_{i1}, x_{i2}, \tilde{x}_k) = 0$ if $\tilde{x}_k = x_{i1} x_{i2}$ and equals 1 or 3 otherwise.

$$h(x_{i1}, x_{i2}, \tilde{x}_k) = 3\tilde{x}_k^2 + x_{i1} x_{i2} - 2x_{i1}\tilde{x}_k - 2x_{i2}\tilde{x}_k \quad (11)$$

QUBO has also been utilised in other transport domains. For example, Dixit and Niu (2023) applied the QUBO formulation to transport network design problems and solved them using QA. In their study, they modelled the link travel time function using the traditional Bureau of Public Roads (BPR) link cost function:

$$t_a(x_a) = t_{f_a}\left(1 + \alpha \left(\frac{x_a}{c_a}\right)^\beta\right) \quad (12)$$

where $t_a$ represents travel time on link $a$, $x_a$ represents Traffic flow on link $a$, $t_{f_a}$ is the free flow travel time on link $a$, $c_a$ is the capacity per lane of link $a$, $\alpha$ and $\beta$ are the attributes of the travel time function. The capacity of link $a$ is expressed as a function of the decision variable $y_a$, which indicates whether the link is selected for capacity expansion.

$$c_a = (l_a + \Delta_a y_a) \tag{13}$$

The link travel time as a function $x_a$ and $y_a$ is expressed as:

$$t_a(x_a, y_a) = t_{fa}\left(1 + \alpha \left(\frac{x_a}{c_a(l_a)}\right)^\beta + \alpha y_a \left[\left(\frac{x_a}{c_a(l_a + \Delta_a)}\right)^\beta - \left(\frac{x_a}{c_a l_a}\right)^\beta\right]\right) \tag{14}$$

Including the slack variable $s_i$, the objective function is formulated as:

$$\begin{aligned}
\min & \sum_a \left(x_a t_{fa}\left(1 + \left(\frac{x_a}{c_a l_a}\right)^\beta\right)\right) \\
& + \sum_a \left(\alpha x_a t_{fa}\left[\left(\frac{x_a}{c_a(l_a + \Delta_a)}\right)^\beta - \left(\frac{x_a}{c_a l_a}\right)^\beta\right] y_a\right) \\
& + \lambda \left((1 - 2N)\sum_a y_a + \sum_{i=1..N}(1 - 2N)s_i \right. \\
& + \sum_{i \in A}\sum_{j \in A\ s.t\ i<j} 2 y_i y_j + \sum_{i=1..|A|}\sum_a 2 s_i y_a \\
& \left. + \sum_{i=1..|A|}\sum_{j=i+1..|A|\ s.t\ i<j} 2 s_i s_j + N^2\right)
\end{aligned} \tag{15}$$

Both QA and gate-based algorithms can solve QUBO problems. While gate-based QCs can solve small-size QUBO problems, QA is more suited to solving large-size QUBO transport optimisation problems as this approach can naturally map binary optimisation problems into qubits. The following sections describe the difference between gate-based and QA paradigms.

### 3. The gate-based Quantum Optimization

#### 3.1 Basic Principles

Gate-based QC operates through a series of quantum gates applied to qubits. This process is analogous to logic gates in classical computing, but it uses qubits and quantum gates applied in a sequence known as a quantum circuit. This approach leverages quantum properties such as superposition and entanglement by manipulating the qubits' states to perform complex calculations. Typically, a qubit in a state of superposition is represented as $|\psi\rangle = \alpha |0\rangle + \beta |1\rangle$ with $\alpha$ and $\beta$ being complex numbers that satisfy the condition $|\alpha|^2 + |\beta|^2 = 1$.

Gate-based QC include several models, each with distinct approaches. The standard quantum circuit model is a common gate-based QC method. Examples of quantum gates include Pauli gates (X, Y, Z gates), Hadamard Gate (H Gate), Controlled Not gate, Toffoli gate, Swap gate, Phase shift gate, and Rotation gates (Copsey et al., 2003; Kwon et al., 2021). An $n$-qubits quantum gate is described by $2^n \times 2^n$ unitary matrices $(U)$. These gates must ensure that the transformation is reversible and that the total probability is preserved (Alsaiyari & Felemban, 2023).

$$\text{X Gate, Bit-flip, Not} \quad \boxed{X} \equiv \begin{bmatrix} 0 & 1 \\ 1 & 0 \end{bmatrix} \begin{bmatrix} \alpha \\ \beta \end{bmatrix} = \beta|0\rangle + \alpha|1\rangle \tag{16}$$

$$\text{Z Gate, Phase-flip} \quad \boxed{Z} \equiv \begin{bmatrix} 1 & 0 \\ 0 & -1 \end{bmatrix} \begin{bmatrix} \alpha \\ \beta \end{bmatrix} = \alpha|0\rangle - \beta|1\rangle \tag{17}$$

$$\text{H Gate, Hadamard} \quad \boxed{H} \equiv \frac{1}{\sqrt{2}}\begin{bmatrix} 1 & 1 \\ 1 & -1 \end{bmatrix} \begin{bmatrix} \alpha \\ \beta \end{bmatrix} = \frac{(\alpha+\beta)|0\rangle + (\alpha-\beta)|1\rangle}{\sqrt{2}} \tag{18}$$

### 3.2 Algorithms

One of the most widely used gate-based quantum algorithms in the transport literature is the Quantum Fourier Transform (QFT) (V. Dixit & Jian, 2022; Heo et al., 2019; Li et al., 2018, 2019). Analogous to the classical Discrete Fourier Transform (DFT), the QFT transforms the amplitudes of a quantum state from the computational basis into a frequency basis. QFT has demonstrated speedups in problems involving hidden periodic structure or frequency analysis such as evaluating the frequency domain of traffic drive cycles, as demonstrated by Dixit and Jian (2022).

In the literature, QFT is implemented on three primary types of gates: (a) the Hadamard Gate, (b) the Controlled Rotation Gate (CROTk), and (c) the Swap Gate. To perform QFT, an $n$-qubit system is initialised, with $N = 2^n$ possible states. The QFT transforms the input qubit states $x_i$ into Fourier coefficients $y_i$, which represent the probabilities of observing specific frequencies in the quantum state.

$$\sum_i x_i |i\rangle \xrightarrow{QFT} \sum_i y_i |i\rangle \tag{19}$$

The QFT is applied to a vector of length $N = 2^n$ represented as $|x\rangle = |x1\rangle \otimes |x2\rangle \otimes \cdots \otimes |xn\rangle$ and $x = 2^{n-1}x_n + \cdots + 2x_1 + x_0$. The Fourier coefficient for each frequency obtained from the QFT represents the probability of observing that frequency:

$$QFT_N|x\rangle = \frac{1}{\sqrt{N}}\sum_y e^{2\pi i \frac{xy}{2^n}}|y\rangle \tag{20}$$

Then a sequence of gates is defined and controlled phase-rotation gates are applied. This process is repeated, with gates are swapped to reverse the order of qubits. As shown in **Figure 1**, QFT utilise the quantum oracle function $U_f$, which processes a superposition of qubits to achieve "quantum parallelism". For the first qubit, $n$ gates are required, consisting of one Hadamard gate and $n-1$ controlled rotation gates. Subsequent qubits progressively require fewer gates, with qubit $k$ using $n-k+1$ gates, followed by $n/2$ swap gates. Then the total number of operations required is $\frac{n^2}{2} + n$. Therefore, the complexity of the QFT is $O(n^2)$ or $O((\ln N)^2)$, representing an exponential improvement compared to the classical Fast Fourier Transform (Nielsen & Chuang, 2010).

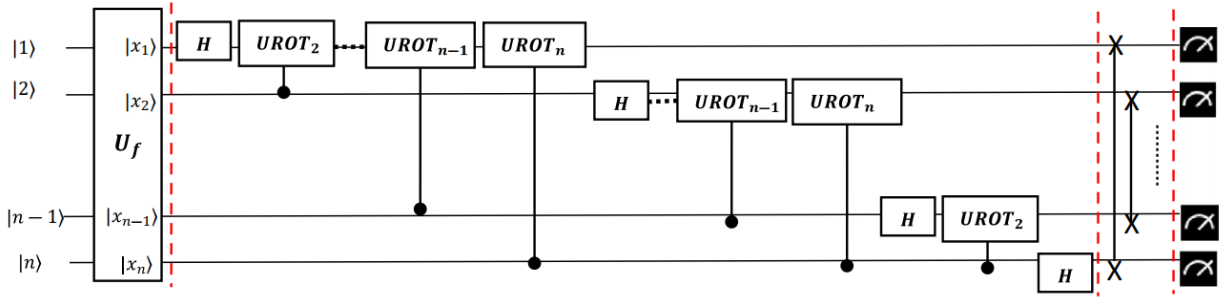

**Figure 1** Circuit for a quantum Fourier transform. (see V. Dixit & Jian, 2022).

Another widely utilised algorithm is Quantum Approximate Optimization Algorithm (QAOA), introduced by Farhi et al.(2000, 2014). QAOA is a hybrid quantum-classical algorithm specifically designed to work on near-term quantum devices, which may have limitations such as a small number of qubits and higher error rates. Additional challenges include the possibility for gradients to become exponentially small in the optimisation space (barren plateaus) and the classical optimisation landscape being NP-hard in general. While QFT is more suited for problems involving frequency analysis, QAOA is more suited for combinatorial optimization problems such as vehicle routing problem (VRP) (Alsaiyari & Felemban, 2023; Azad et al., 2023; Gautam & Ahn, 2023; Mohanty et al., 2023), and job scheduling such as bike sharing rebalancing problem (Harikrishnakumar & Nannapaneni, 2021).

QAOA operates on a multi-step process. First, the optimization problem is encoded, often in the form of QUBO, or a Hamiltonian cost function. This step translates the problem into a format that a quantum computer can handle. The algorithm uses a sequence of quantum gates applied to each qubit, putting them into a superposition of states. QAOA alternates between two quantum operations, as demonstrated in **Figure 2**. The first operator is called 'phase separation" in which the cost Hamiltonian layer is expressed as the unitary operator $U(H_{C,\gamma})$ with parameter $\gamma$:

$$U(H_C, \gamma) = e^{-i\gamma H_c} \qquad (21)$$

The second operator is mixing operator or mixer Hamiltonian layer, which is expressed as a unitary operator $U(H_M, \beta)$ with parameter $\beta$:

$$U(H_M, \beta) = e^{-i\beta H_M} \qquad (22)$$

These two operators are applied for a certain number of layers. For $p$ layers, the angle dependent quantum state can be expressed as alternating applications of operations (22) and (23) on the initial state. Mathematically, it is represented as:

$$|\gamma, \beta\rangle = U(H_M, \beta_p) U(H_C, \gamma_p) \cdots U(H_M, \beta_0) U(H_C, \gamma_0) |\psi_0\rangle \qquad (23)$$

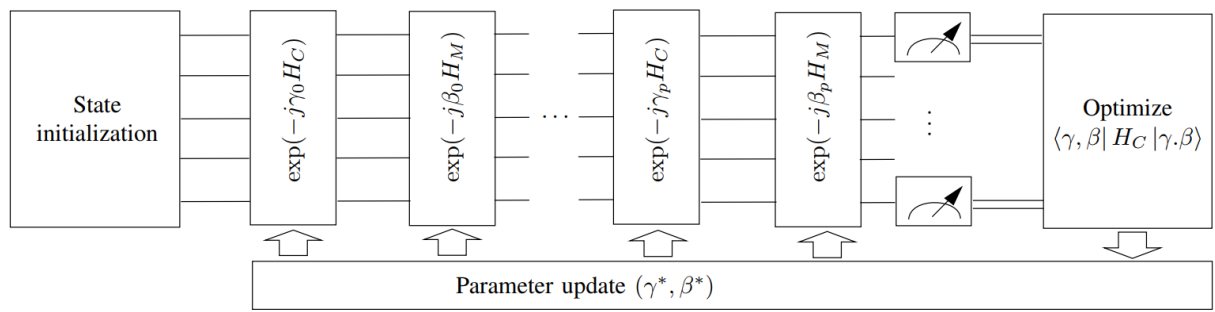

**Figure 2** The general workflow of QAQA. (see Sinaga et al., 2023).

In each layer, the parameters $\vec{\gamma}$ and $\vec{\beta}$ are tuned to maximise or minimise the probability of measuring the quantum state corresponding to the optimal or near optimal. These parameters are crucial for determining the behaviour of the quantum system under simulation. The goal is to adjust these parameters so that the resulting quantum state minimises the Hamiltonian cost function, thus providing an optimal solution to the original problem. Parameter adjustment is done through a classical optimization algorithm, creating a powerful synergy between quantum and classical computing techniques. This hybrid approach allows QAOA to perform effectively even on quantum devices that are not yet fault-tolerant, making it a promising approach for early applications of quantum computing in solving practical problems.

### 3.3 Applications and Implementation

Gate-based QC theoretically can solve any computational problem, including but not limited to optimization, simulation, and machine learning. Its implementation requires qubits that can be precisely controlled with quantum gates and maintained in a coherent state throughout the computation. Table 1 provides detailed information about the quantum solver, decision variables, objective, constraints and benchmarking methods.

The hardware implementation for gate-based QC includes superconducting qubits, trapped ions, and photonic systems. Despite their potential, gate-based QC systems face significant challenges. Quantum decoherence, where qubits lose their quantum state due to environmental disturbances, is a major issue. Researchers are working on improving qubit control and developing error correction algorithms to mitigate these challenges (Leymann & Barzen, 2020).

Currently, the gate-based QC is still in early stages of technology maturity, with practical applications limited to small-scale quantum processors. As technology advances, larger and more reliable quantum systems are expected to become available, potentially revolutionising various fields by solving complex problems more efficiently than classical computers.

Table 1 Gate-based QC application in transportation studies

| Reference | Quantum solver | Decision variables | Objective | Constraints | Formulation | Benchmark methods | Problem |
|---|---|---|---|---|---|---|---|
| **Dixit and Jian (2022)** | IBM-Q16 Melbourne | Binary | Not applicable | Not applicable | QFT | Classical Fourier transform | Drive cycle frequency estimation |
| **Harikrishnakumar and Nannapaneni (2021)** | IBM-Qiskit | Binary | Min total travel distance | Supply and demand balance constraint, parking lot cycle constraint, Single return constraint, one-time visit constraint | QUBO and Hamiltonian | Not given | Bike sharing system rebalancing problem |
| **Alsaiyari and Felemban (2023)** | IBM Qiskit | Binary | Min total travel distance | Single visit constraint, depot connection constraint | Hamiltonians | CPLEX | Vehicle routing problem |
| **Harikrishnakumar and Nannapaneni (2023)** | IBM Qiskit | Binary | Not applicable | Not applicable | Quantum Bayesian Network | Netica | Bike demand prediction problem |
| **Azad et al. (2023)** | IBM Qiskit | Binary | Min total travel distance | Single visit constraint, routing sequencing constraint | QUBO and Hamiltonians | CPLEX | Vehicle routing problem |
| **Mohanty et al. (2023)** | IBM Qiskit | Binary | Min total travel cost | Single visit constraint, routing sequencing constraint | QUBO and Hamiltonians | Not given | Vehicle routing problem |
| **Sinaga et al. (2023)** | Not given | Binary | Min total travel cost | Exclusive visit constraint, one-time visit constraint | Hamiltonians | Brute force | Travel salesman problem |

## 4. The quantum annealing (QA) Optimization

### 4.1 Basic Principles

QA, first introduced by Kadowaki and Nishimori (Kadowaki, 2002; Kadowaki & Nishimori, 1998), is as a quantum optimization technique inspired by classical simulated annealing. In QA, the optimization process begins with qubits initialised in a superposition state. The annealing process involves slowly decreasing a quantum parameter while increasing the problem Hamiltonian's influence, allowing the system to settle into a low-energy configuration that represents a solution. The system then undergo a gradual evolution governed by a time-dependent Hamiltonian formulation as follows:

$$H_{QA}/h = \underbrace{-A(t/t_f)\sum_i \sigma_i^X}_{Initial\ Hamiltonian} + \underbrace{B(t/t_f)\left(\sum_i h_i \sigma_i^Z + \sum_{i>j} J_{ij}\sigma_i^Z \sigma_j^Z\right)}_{Final\ Hamiltonian} \quad (24)$$

where $t_f$ is the annealing time, $\sigma_i^X$ and $\sigma_i^Z$ are the Pauli matrices acting on qubit $i$, and $h_i$ and $J_{ij}$ are the qubit biases and coupling strengths, respectively. The quantum annealing starts with the ground state of the initial Hamiltonian, where all qubits are in a superposition of |0⟩ and |1⟩. As the system evolves according to a predefined annealing schedule, governed by time-dependent functions $A(t/tf)$ and $B(t/tf)$, the Hamiltonian gradually shifts from its initial form to the final state, which encodes the solution to the optimization problem. $A(t/tf)$ represents the transverse energy that facilitates quantum tunnelling between states, while $B(t/tf)$ corresponds to the energy applied to the problem Hamiltonian. These energy scales, $A(t/tf)$ and $B(t/tf)$, dynamically adjust throughout the quantum annealing process, reflecting the gradual shift from a quantum state dominated by quantum fluctuations to a classical state focused on solving the optimization problem.

### 4.2 Algorithms

D-Wave, one of the pioneering QA technologies, typically utilises QUBO model formulation or Ising Hamiltonian model. Ising Hamiltonian is isomorphic to QUBO Problems, and are of the form:

$$\text{Obj} := x^T Q_x \quad (25)$$

where $x$ is a vector of $N$ binary variables and $Q$ is a $N \times N$ matrix representing the coefficients of the quadratic terms. The diagonal terms of $Q$ are mapped to $h_i$ and the cross terms are mapped to $J_{ij}$ in the final Hamiltonian. The process of embedding involves mapping a problem onto the physical qubits and couplers available in the D-Wave quantum processor. This step is crucial for translating real-world problems into a form that can be processed by the quantum annealer.

### 4.3 Applications and Implementation

QA is implemented using superconducting qubits, such as those in D-Wave systems, which are designed to behave in a way that mimics the quantum annealing process. Unlike gate-based approach, QA does not require gate-based controls and swapping. With improvements in error correction algorithms(Amin et al., 2023; Pearson et al., 2019) and the growing number of qubits (D-Wave, 2024c), QA technology is more mature and scalable, enabling the use of a larger number of qubits compared to gate-based quantum computing.

QA is very well suited to explore large combinatorial space quickly. For example, solving the VRP using gate-based QC is constrained by the limited number of qubits, typically fewer than 100 , which necessitates the design of specific quantum circuits. In contrast, QA can handle larger-scale computational problems more efficiently. This advantage is evident in its application to large instances of VRP (Ajagekar & You, 2022; Feld et al., 2019; Harikrishnakumar, Nannapaneni, et al., 2020; Osaba et al., 2021; Papalitsas et al., 2019).

QA has also been successfully applied to various other transport-related optimization problems, including railway rescheduling (Domino et al., 2022), traffic signals control (Hussain et al., 2020), traffic flow optimization problem (Neukart et al., 2017), and air traffic management (Stollenwerk et al., 2020). Table 2 provides a detailed information about the quantum solver, decision variables, objective, constraints and benchmarking methods.

Table 2 QA in transportation

| Reference | Quantum solver | Decision variables | Objective | Constraints | Formulation | Benchmark methods | Problem |
|---|---|---|---|---|---|---|---|
| **Suen et al. (2021)** | Fujitsu's quantum inspired Digital Annealer | Binary | Min total travel time | Flow conservation constraint, Once-visit constraint | QUBO | CPLEX, Nearest-neighbour heuristics | Vehicle sharing problem |
| **Papalitsas et al. (2019)** | D-Wave 2000Q QPU | Binary | Min total travel cost, min time to return to the depot | Feasible tour constraint; Time window constraint | QUBO | Not given | Traveling salesman problem, scheduling problem |
| **Ajagekar et al (2020)** | D-Wave 2000Q QPU | Binary | Min ratio of travel costs to resources spent | Once-visit constraint, Routing sequencing constraint | QUBO with QC-IQFP parametric method | Baron 19, Bonmin 15 | Vehicle routing problem |
| **Sales and Araos (2023)** | Amazon Bracket | Binary | Min total travel cost | Once-visit constraint, Routing sequencing constraint, capacity constraint | QUBO | K-Medoids algorithm and OR-Tools | Vehicle routing problem |
| **Domino et al. (2022)** | D-Wave Advantage, D-Wave hybrid solver | Binary | Min weighted additional delay | Headway constraint, rolling stock circulation constraint, track occupancy constraint | QUBO, HOBO | None | Railway rescheduling |
| **Feld et al.(2019)** | D-Wave 2000Q | Binary | Min total travel distance | Routing sequencing constraint, capacity constraint, | QUBO | Best known solution* | Capacitated vehicle routing problem |
| **Hussain et al. (2020)** | D-Wave 2000Q_5 | Binary | Max traffic flow, Min | Mode selection constraint, | QUBO | QBSolv tabu solver | Traffic signals control |

| | | | total wasted time | Intersection coordination constraint | | | |
|---|---|---|---|---|---|---|---|
| **Harikrishnakumar et al. (2020)** | None | Binary | Min total travel distance | Once-visit constraint, Routing sequencing constraint, capacity constraint | QUBO | None | Multi-depot capacitated vehicle routing problem |
| **Neukart et al. (2017)** | D-Wave 2X QPU | Binary | Min total congestion | Car route constraint | QUBO | Random assignment and original assignment | Traffic flow optimization problem |
| **Jain (2021)** | D-Wave's Advantage 1.1 QPU | Binary | Min total travel cost | Once-visit constraint | QUBO | Heuristic and brute-force | Traveling salesman problem |
| **Stollenwerk et al. (2020)** | D-Wave 2X, D-Wave 2000Q | Binary | Min total flights arrival delay | Conflict constraint | QUBO | Isoenergetic Cluster Method | Air traffic management |
| **Niu et al. (2024)** | D-Wave hybrid solver | Binary, Integer | Min total travel cost | Flow conservation constraint, capacity constraint | MIP | CPLEX | Multicommodity network flow problem |
| **Dixit and Niu (2023)** | D-Wave Advantage™ quantum computer and hybrid solver | Binary | Min total system travel time | Flow conservation constraint | QUBO | Tabu search | Transport network design problem |
| **Dixit et al. (2023)** | D-Wave hybrid solver | Binary | Min expected travel cost | Flow conservation constraint, routing sequencing constraint | MIP | CPLEX, Naïve algorithm | Scenario-based stochastic time-dependent shortest path Routing problem |
| **Osaba et al. (2021)** | D-Wave Advantage system 1.1 | Binary | Min total travel distance | Routing sequencing constraint, capacity constraint, | QUBO | Not given | Traveling salesman problem |
| **Xu et al. (2023)** | Kaiwu-SDK, Coherent Ising machine (CIM) quantum simulator | Binary | Min the total train travel time costs, Min the total train departure delay at the departure station, Min the number of train stops | Train stop limit constraint, station service frequency constraint, train stopping time constraint, flow balancing constraint, departure time window constraint, station stopping capacity constraint, train safety interval constraint, interval crossing constraint, maintenance gap constraint | QUBO | Gurobi, simulated annealing | High-speed train timetabling problem |

## 5. The quantum machine learning (QML)

### 5.1 Basic Principles

QML integrates QC with traditional machine learning (ML) techniques, harnessing quantum superposition, entanglement, and interference to handle high-dimensional data, more effectively than classical computers. Unlike classical ML, QML involves encoding classical data into quantum states. This process typically encompasses three principal elements: state encoding, parameterized quantum circuits (PQC), and measurement (Park et al., 2024). Due to superposition, a single qubit can represent multiple states simultaneously, enabling QC to potentially process high-dimensional data classification or pattern prediction with enhanced efficiently.

### 5.2 Algorithms

Several algorithms have been developed for QML, including Quantum Support Vector Machine (QSVM), Quantum Neural Network (QNN), Quantum Graph Convolutional Neural Network (QGCNN), Quantum Deep Learning (QDL), and Quantum Bayesian Network (QBN).

For example, Qu et al. (2023) developed a QGCCN to predict the traffic congestion. They transformed classical traffic data (e.g. $(x_1, \ldots, x_d)$) to quantum state $|\varphi\rangle_{(1,\ldots,d)}$, parameterised quantum gates $U(\vec{\theta}) = U_L(\theta_L) \ldots U_1(\theta_1)$, and utilised quantum walks. **Figure 3** illustrates the framework of their temporal-spatial QGCNN, which leverages the Schrödinger approach to analyse temporal features and captures spatial features of the traffic network. The classical traffic data is transformed into quantum data and the Schrödinger approach is utilized for time dimension analysis, yielding a closed-form solution that reveals temporal features. Subsequently, a QGCNN is developed to capture the spatial features of the traffic network, leveraging the temporal information. The prediction results are finally obtained through the measurement operation $\hat{Z}$.

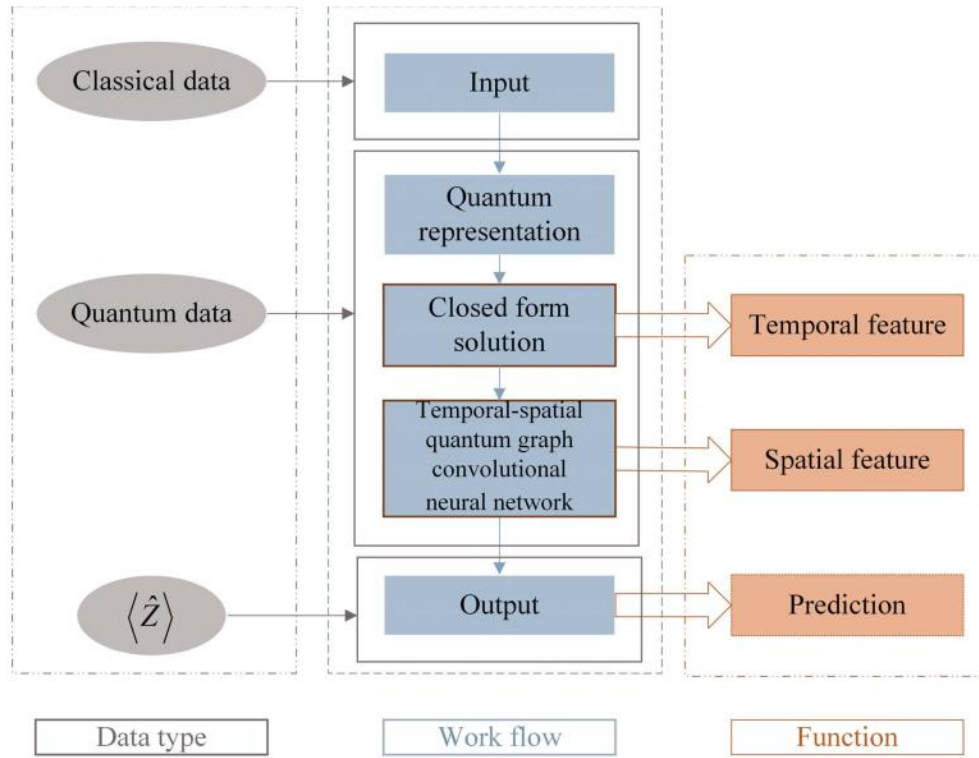

**Figure 3** The framework of the temporal-spatial quantum graph convolutional neural network based on Schrödinger approach. (see Qu et al., 2023).

Majumder et al. (2021) designed hybrid classical-quantum deep learning models for classifying traffic images in autonomous vehicles under adversarial conditions. Their model involves three phases: (i) data embedding phase; (ii) variational circuit phase; and (iii) quantum measurement. In the data embedding phase, classical data is embedded into quantum states using qubit quantum gates such as Hadamard gate, and Rotational Y gates. The variation circuit phase involves parameterized quantum circuits combining single-qubit gates like CNOT, CZ, and CRX. The measurement layer reads the qubit's states, converting them to classical data using X, Y, and Z basis.

Harikrishnakumar and Nannapaneni (2023) developed a quantum Bayesian network to forecast bike sharing demand.

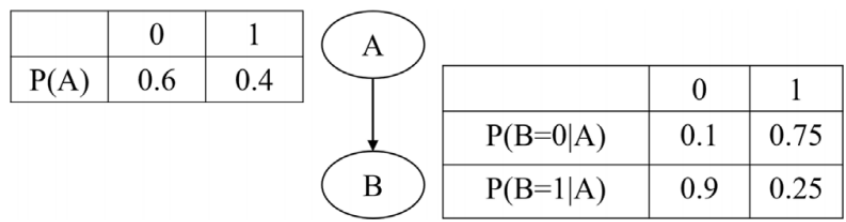

**Figure 4** Illustration of Bayesian network. (see Harikrishnakumar, Borujeni, et al., 2020).

As shown in Figure 4, the Bayesian network nodes correspond to qubit state $|0\rangle$ and $|1\rangle$. The parent node

A is represented with a single qubit gate, and its probability is determined by the probability amplitudes of states $|0\rangle$ and $|1\rangle$, which are adjusted using a controlled rotation gate with a specified rotation. The probabilities of the child node B are conditional on the values of the parent node A. Since A can take two values ($A = 0$ or $A = 1$), two rotation angles are used to represent the probabilities of B corresponding to each value of A.

Haboury et al. (2023) proposed a supervised hybrid quantum neural network (HQNN) that mimics the node-wise Dijkstra's algorithm in a dynamic earthquake emergency evacuation scenario. The model uses feature-wise linear modulation (FiLM) neural networks to enhance adaptability to dynamic changes in the environment. The quantum network compromises seven qubits: five encode path features and neighbour data, two process earthquake coordinates. After entangling these qubits and adding a trainable layer, the main qubits are measured in the Z basis. The HQNN model demonstrates enhanced performance, achieving an average accuracy of 94% and predicting better paths than the Dijkstra algorithm in 25% of cases. The reason is that Node-wise Dijkstra makes real-time travel decisions that can be sub-optimal if conditions change. The HQNN model, in contrast, adapts to dynamic graphs for more robust choices.

### 5.3 Applications and Implementation

QML can be implemented in both gate-based and QA systems for a range of applications, including clustering, classification (Majumder et al., 2021), and prediction of transport and traffic data (Haboury et al., 2023; Harikrishnakumar & Nannapaneni, 2023; Martín-Guerrero & Lamata, 2022; Park et al., 2024; Qu et al., 2023; Schuld et al., 2015). Table 3 provides a detailed information about the quantum solver, decision variables, objective, constraints and benchmarking methods.

Table 3 QML studies in transportation

| Reference | Quantum solver | Benchmark methods | Problem |
|---|---|---|---|
| **Haboury et al. (2023)** | IonQ Aria 1 | Classical neural network | Emergency escape routing problem |
| **Park et al. (2024)** | Not given | Not given | Autonomous mobility cooperation |
| **Yamany et al. (2023)** | Not given | Particle swarm optimization, Hill climbing, Random search | Adversarial attack defending |
| **Majumder et al. (2021)** | PennyLane (2022) | Quantum deep learning– Pre-trained Resnet18 model | Autonomous vehicle traffic image classification under adversarial attack |

## 6. The State of The Art in the QC and Transport Studies

Emerging quantum technologies like the Coherent Ising Machine (CIM) are applied to niche areas such as high-speed train timetabling, while QML is used for dynamic decision-making and real-time tasks like emergency escape routing and autonomous vehicle image classification. Table 4 reflects quantum computing's growing potential to solve large-scale combinatorial optimization problems, with QA offering scalability possibilities, while gate-based QC and QML provide adaptability for more complex prediction and learning tasks in transport data analytics.

In addition to the typical QC problem formulations mentioned above, other recent advancements in hybrid formulations offer new solutions to NP-hard transportation problems. For example, Dixit et al. (2023) proposed a quadratic constrained binary optimization problem for the scenario-based stochastic and time-dependent routing problem. This was solved using a hybrid constrained quadratic model (CQM) solver provided by D-Wave systems. In their study, QA was benchmarked for different sizes of transport networks and achieved computational speeds up to 11 times faster than Tabu Search for large-scale networks.

Similarly, Harikrishnakumar et al. (2020) extended the application of QC to the multi-depot capacitated VRP and its variant. They formulated the vehicle and depot capacity constraints, as follows:

$$\sum_i \sum_j q_i x_{ijk} + \sum_i q_i \eta_{ik} \leq Q_k \qquad (26)$$

$$\sum_k \gamma_{kd} \left( \sum_i \sum_j q_i x_{ijk} + \sum_i q_i \eta_{ik} \right) \leq V_d \qquad (27)$$

where $q_i$ is the demand at location $i$, $\eta_{ik}$ is a binary variable that equals 1 if location $i$ is the last stop served by vehicle $k$ before it returns to its depot, $Q_k$ is the capacity of vehicle $k$, $\gamma_{kd}$ is the binary variable that equals 1 if vehicle $k$ belongs to depot $d$ and $V_d$ is the capacity of depot $d$. When vehicles are in service and new customer requests arise, their current locations become the starting points for rerouting. In the static approach, vehicles start from depots. Rerouting considers the remaining capacities of both vehicles and depots, accounting for already-served locations. The corresponding vehicle capacity and depot capacity constrains are formulated as:

$$\sum_{i \in \Theta w} \sum_{j \in \Theta w} q_i x_{ijk} + \sum_{i \in \Theta w} q_i \eta_{ik} \leq Q_k - \sum_{w \in \Gamma w} q_w \gamma_{kw} \qquad (28)$$

where $\Gamma w$ and $\Theta w$ represent the sets of locations already served and yet to be served, respectively. $\gamma_{kw} = 1$ if vehicle $k$ served a location $w$ in $\Gamma w$.

$$\sum_k \gamma_{kd}(\sum_{i\epsilon\Theta w}\sum_{j\epsilon\Theta w}q_i x_{ijk} + \sum_{i\epsilon\Theta w}q_i\eta_{ik}) \leq V_d - \sum_k\sum_{w\epsilon\Gamma w}q_w\gamma_{kw}\gamma_{kd} \qquad (29)$$

In addition to these advancements, Feld et al. (2019) proposed a two-phase heuristic approach for the capacitated vehicle routing problem (CVRP), which splits the problem into clustering phase and routing phases, as demonstrated in **Figure 5**. This approach explores different mappings for subproblems, employing QUBO and classical methods. The geometric centre $CC\ (m_k)$ of cluster $m_k$ is calculated as

$$CC(m_k) = \sum_i \frac{v_i^x}{n}, \sum_i \frac{v_i^y}{n} \qquad (30)$$

where $v_i^x$ and $v_i^y$ represent the $x$ and $y$ coordinates of customer $v_i$, and $n$ denotes the total number of customers in cluster $m_k$. The preliminary study investigates mapping the subproblems in three ways: as two separate QUBOs, as a single QUBO, and by combining classic methods for the clustering phase with a QUBO approach for the routing phase. The results indicate that this hybrid method yields the best solutions.

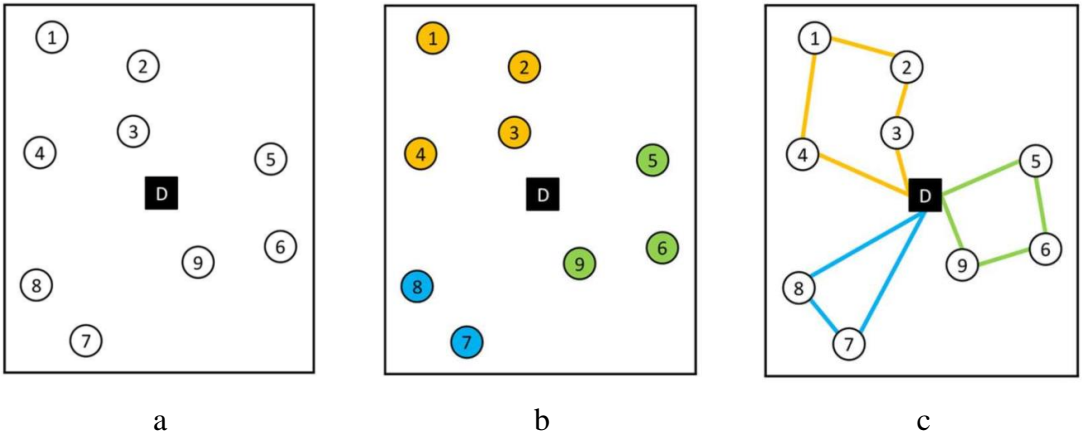

a b c

**Figure 5** The two-phase heuristic method. (a) Starting with 9 customers and 1 depot. (b) Clustering creates three clusters. (c) shortest paths in each cluster. (see Feld et al. 2019).

The evolving landscape of QC demonstrates significant potential for optimising combinatorial problems in transportation. As shown in Table 1, Table 2 and Table 3, key solvers include D-Wave's QPUs, Fujitsu's Digital Annealer, and Amazon Bracket, which work with binary decision variables and aim to minimise costs, travel distances, or delays. The QUBO model is widely used across the studies, where typical constraints include flow conservation, capacity limits, and routing sequences to ensure feasible solutions. In addition, Mixed Integer Programming (MIP) is used in hybrid approaches, combining classical and quantum solvers for complex problems like multicommodity network flows and scenario-based stochastic time-dependent routing.

Gate-based quantum computing, particularly using IBM Qiskit, has been applied to problems like bike-sharing system rebalancing, VRP, and demand prediction, using algorithms like QAOA and quantum Bayesian networks. Emerging fields such as QML are being used for emergency escape routing and adversarial attack defence. For benchmarking, classical solvers such as CPLEX, simulated annealing, and tabu search are frequently used to compare the performance of quantum solutions, with some hybrid methods demonstrating significant speed-ups, especially in large-scale networks. The problems addressed range from classical VRP to more specialised challenges like high-speed train timetabling and autonomous vehicle operations, showcasing the growing applicability of QC in solving complex transportation problems.

**Table 4** Literature review for the applications of QC in transportation system

| Transportation Problem | Formulation | QC Paradigm | Reference |
|---|---|---|---|
| Vehicle Routing Problem | QUBO, | QA | (Ajagekar et al., 2020; Feld et al., 2019; Harikrishnakumar, Nannapaneni, et al., 2020; Jain, 2021; Osaba et al., 2021; Papalitsas et al., 2019; Sales & Araos, 2023; Suen et al., 2021) |
| | | Gate-based QC, using QAOA | (Alsaiyari & Felemban, 2023; Sinaga et al., 2023); (Azad et al., 2023; Gautam & Ahn, 2023; Mohanty et al., 2023) |
| Railway rescheduling | QUBO, HOBO | QA | (Domino et al., 2022) |
| Traffic signals control | QUBO | QA | (Hussain et al., 2020) |
| Traffic flow optimization problem | QUBO | QA | (Neukart et al., 2017) |
| Air traffic management | QUBO | QA | (Stollenwerk et al., 2020) |
| Transport network design problem | QUBO | QA | (V. V. Dixit & Niu, 2023) |
| High-speed train timetabling problem | QUBO | Coherent Ising machine (CIM) quantum simulator | (Xu et al., 2023) |
| Multicommodity network flow problem | MIP | QA | (Niu et al., 2024) |
| Scenario-based stochastic time-dependent shortest path Routing problem | MIP | QA | (V. V. Dixit et al., 2023) |
| Drive cycle frequency estimation | QFT | Gate-based QC, using QAOA | (V. Dixit & Jian, 2022) |
| Bike sharing system rebalancing problem | QUBO | Gate-based QC, using QAOA | (Harikrishnakumar & Nannapaneni, 2021) |
| Bike demand prediction problem | Quantum Bayesian Network | Gate-based QC, using QAOA | (Harikrishnakumar & Nannapaneni, 2023) |
| Emergency escape routing problem | Not applicable | QML | (Haboury et al., 2023) |
| Autonomous mobility cooperation | Not applicable | QML | (Park et al., 2024) |
| Adversarial attack defending | Not applicable | QML | (Yamany et al., 2023) |
| Autonomous vehicle traffic image classification under adversarial attack | Not applicable | QML | (Majumder et al., 2021) |

## 7. Conclusion

Recent advancements in QC have spurred significant interest in its application to transportation systems. Notably, with IBM's development of the Condor processor, featuring 1,121 superconducting qubit using cross-resonance gate technology (IBM, 2023) and D-Wave's system now boasting over 5000 qubits (D-Wave, 2024b), alongside innovative structural designs (D-Wave, 2024a), the potential of QC to tackle complex optimization problems is becoming increasingly evident.

Gate based quantum computers have been theoretically shown to offer substantial speedup for specific classes of problems. For instance, they provide quadratic speedup for search problems (Grover's algorithm) and exponential speedup for solving system of linear equations (HHB) and Prime Factorization (Shor's algorithm). This has shifted some problems, such as prime factorization, from the NP Class of Complexity to Bounded-error Quantum Polynomial (BQP) time complexity. However, it is important to note that efficient quantum algorithms for NP-Complete problems have yet to be discovered. Approximation quantum algorithms like QAOA provide time efficient algorithms, but do not guarantee efficient solutions. Furthermore, the current limitations in gate-based systems – including the number of qubits and errors caused by decoherence – restrict their practical applications to NP-complete problems.

Where gate-based systems have been limited in its application to NP-complete problems, QA approaches have been highly successful in solving QUBO problems. While QA operates somewhat as a black box, relying on the physics of annealing to solve the embedded problem, strategies such as problem decomposition and modified Lagrange multipliers can improve its performance. These methods, coupled with hybrid approaches that incorporates heuristic algorithms, are poised to maximize the potential of quantum resources. This approach is particularly crucial in fields requiring robust computational solutions such as transportation systems.

QML, particularly with gate-based QC, currently faces scalability issues due to the limited number of qubits, which hampers its applicability to real-world problems. However, QML relying on QA offer a more promising approach, potentially overcoming these barriers. QML could also be instrumental in developing intelligent control systems for autonomous vehicles, while the integration of quantum information theory (Bennett & DiVincenzo, 2000) may enhance cybersecurity measures for emerging technologies such as autonomous vehicles and flying cars.

Looking forward, the emergence of QC methods holds the promise of developing efficient transportation systems in the near future. These advancements suggest a paradigm shift where QC could play a critical role in addressing complex optimisation problems and enhancing system efficiencies across various industries.

**Acknowledgments**

This research did not receive any support from any funding agencies.

**Author contributions**

The authors declare no competing interests.

## 8. References

Ajagekar, A., Humble, T., & You, F. (2020). Quantum computing based hybrid solution strategies for large-scale discrete-continuous optimization problems. *Computers & Chemical Engineering*, *132*, 106630. https://doi.org/10.1016/j.compchemeng.2019.106630

Ajagekar, A., & You, F. (2019). Quantum computing for energy systems optimization: Challenges and opportunities. *Energy*, *179*, 76–89. https://doi.org/10.1016/j.energy.2019.04.186

Ajagekar, A., & You, F. (2022). Quantum computing and quantum artificial intelligence for renewable and sustainable energy: A emerging prospect towards climate neutrality. *Renewable and Sustainable Energy Reviews*, *165*, 112493. https://doi.org/10.1016/j.rser.2022.112493

Alsaiyari, M., & Felemban, M. (2023). Variational Quantum Algorithms for Solving Vehicle Routing Problem. *2023 International Conference on Smart Computing and Application (ICSCA)*, 1–4. https://doi.org/10.1109/ICSCA57840.2023.10087522

Amin, M. H., King, A. D., Raymond, J., Harris, R., Bernoudy, W., Berkley, A. J., Boothby, K., Smirnov, A., Altomare, F., Babcock, M., Baron, C., Connor, J., Dehn, M., Enderud, C., Hoskinson, E., Huang, S., Johnson, M. W., Ladizinsky, E., Lanting, T., … Franz, M. (2023). *Quantum error mitigation in quantum annealing* (No. arXiv:2311.01306). arXiv. http://arxiv.org/abs/2311.01306

Arute, F., Arya, K., Babbush, R., Bacon, D., Bardin, J. C., Barends, R., Biswas, R., Boixo, S., Brandao, F. G., & Buell, D. A. (2019). Quantum supremacy using a programmable superconducting processor. *Nature*, *574*(7779), 505–510.

Azad, U., Behera, B. K., Ahmed, E. A., Panigrahi, P. K., & Farouk, A. (2023). Solving Vehicle Routing Problem Using Quantum Approximate Optimization Algorithm. *IEEE Transactions on Intelligent Transportation Systems*, *24*(7), 7564–7573. https://doi.org/10.1109/TITS.2022.3172241

Benioff, P. (1980). The computer as a physical system: A microscopic quantum mechanical Hamiltonian model of computers as represented by Turing machines. *Journal of Statistical Physics*, *22*(5), 563–591. https://doi.org/10.1007/BF01011339

Bennett, C. H., & DiVincenzo, D. P. (2000). Quantum information and computation. *Nature*, *404*(6775), 247–255.

Bergholm, V., Izaac, J., Schuld, M., Gogolin, C., Ahmed, S., Ajith, V., Alam, M. S., Alonso-Linaje, G., AkashNarayanan, B., Asadi, A., Arrazola, J. M., Azad, U., Banning, S., Blank, C., Bromley, T. R., Cordier, B. A., Ceroni, J., Delgado, A., Di Matteo, O., … Killoran, N. (2022). *PennyLane: Automatic differentiation of hybrid quantum-classical computations* (No. arXiv:1811.04968). arXiv. https://doi.org/10.48550/arXiv.1811.04968

Bourreau, E., Fleury, G., & Lacomme, P. (2022). *Mixer Hamiltonian with QAOA for Max k-coloring: Numerical evaluations* (No. arXiv:2207.11520). arXiv. https://doi.org/10.48550/arXiv.2207.11520

Bova, F., Goldfarb, A., & Melko, R. G. (2021). Commercial applications of quantum computing. *EPJ Quantum Technology*, *8*(1), 2. https://doi.org/10.1140/epjqt/s40507-021-00091-1

Chen, Z.-Y., Xue, C., Chen, S.-M., Lu, B.-H., Wu, Y.-C., Ding, J.-C., Huang, S.-H., & Guo, G.-P. (2022). Quantum approach to accelerate finite volume method on steady computational fluid dynamics problems. *Quantum Information Processing*, *21*(4), 137.

Cooper, C. H. V. (2022). Exploring Potential Applications of Quantum Computing in Transportation Modelling. *IEEE Transactions on Intelligent Transportation Systems*, *23*(9), 14712–14720. https://doi.org/10.1109/TITS.2021.3132161

Copsey, D., Oskin, M., Impens, F., Metodiev, T., Cross, A., Chong, F. T., Chuang, I. L., & Kubiatowicz, J. (2003). Toward a scalable, silicon-based quantum computing architecture. *IEEE Journal of Selected Topics in Quantum Electronics*, *9*(6), 1552–1569. https://doi.org/10.1109/JSTQE.2003.820922

Dixit, V., & Jian, S. (2022). Quantum Fourier transform to estimate drive cycles. *Scientific Reports*, *12*(1), 654.

Dixit, V. V., & Niu, C. (2023). Quantum computing for transport network design problems. *Scientific Reports*, *13*(1), 12267. https://doi.org/10.1038/s41598-023-38787-2

Dixit, V. V., Niu, C., Rey, D., Waller, S. T., & Levin, M. W. (2023). Quantum Computing to Solve Scenario-Based Stochastic Time-Dependent Shortest Path Routing. *Transportation Letters*, 1–11. https://doi.org/10.1080/19427867.2023.2238461

Domino, K., Kundu, A., Salehi, Ö., & Krawiec, K. (2022). Quadratic and higher-order unconstrained binary optimization of railway rescheduling for quantum computing. *Quantum Information Processing*, *21*(9), 337. https://doi.org/10.1007/s11128-022-03670-y

D-Wave. (2024a). *D-Wave QPU Architecture: Topologies—D-Wave System Documentation documentation*. https://docs.dwavesys.com/docs/latest/c_gs_4.html

D-Wave. (2024b). *The Advantage™ Quantum Computer | D-Wave*. https://www.dwavesys.com/solutions-and-products/systems/

D-Wave. (2024c, November 6). *D-Wave Achieves Significant Milestone with Calibration of 4,400+ Qubit Advantage2 Processor*. https://www.dwavesys.com/company/newsroom/press-release/d-wave-achieves-significant-milestone-with-calibration-of-4-400-qubit-advantage2-processor/

Farhi, E., Goldstone, J., & Gutmann, S. (2014). *A Quantum Approximate Optimization Algorithm* (No. arXiv:1411.4028). arXiv. https://doi.org/10.48550/arXiv.1411.4028

Farhi, E., Goldstone, J., Gutmann, S., & Sipser, M. (2000). *Quantum Computation by Adiabatic Evolution* (No. arXiv:quant-ph/0001106). arXiv. https://doi.org/10.48550/arXiv.quant-ph/0001106

Feld, S., Roch, C., Gabor, T., Seidel, C., Neukart, F., Galter, I., Mauerer, W., & Linnhoff-Popien, C. (2019). A Hybrid Solution Method for the Capacitated Vehicle Routing Problem Using a Quantum Annealer. *Frontiers in ICT*, *6*, 13. https://doi.org/10.3389/fict.2019.00013

Feynman, R. P. (2018). Simulating physics with computers. In *Feynman and computation* (pp. 133–153). cRc Press. https://www.taylorfrancis.com/chapters/edit/10.1201/9780429500459-11/simulating-physics-computers-richard-feynman

Gautam, K., & Ahn, C. W. (2023). Quantum Path Integral Approach for Vehicle Routing Optimization With Limited Qubit. *IEEE Transactions on Intelligent Transportation Systems*, 1–15. https://doi.org/10.1109/TITS.2023.3327157

Grover, L. K. (1996). A fast quantum mechanical algorithm for database search. *Proceedings of the Twenty-Eighth Annual ACM Symposium on Theory of Computing*, 212–219.

Haboury, N., Kordzanganeh, M., Schmitt, S., Joshi, A., Tokarev, I., Abdallah, L., Kurkin, A., Kyriacou, B., & Melnikov, A. (2023). *A supervised hybrid quantum machine learning solution to the emergency escape routing problem* (No. arXiv:2307.15682). arXiv. https://doi.org/10.48550/arXiv.2307.15682

Harikrishnakumar, R., Borujeni, S. E., Dand, A., & Nannapaneni, S. (2020). A Quantum Bayesian Approach for Bike Sharing Demand Prediction. *2020 IEEE International Conference on Big Data (Big Data)*, 2401–2409. https://doi.org/10.1109/BigData50022.2020.9378271

Harikrishnakumar, R., & Nannapaneni, S. (2021). Smart Rebalancing for Bike

Sharing Systems using Quantum Approximate Optimization Algorithm. *2021 IEEE International Intelligent Transportation Systems Conference (ITSC)*, 2257–2263. https://doi.org/10.1109/ITSC48978.2021.9564714

Harikrishnakumar, R., & Nannapaneni, S. (2023). Forecasting Bike Sharing Demand Using Quantum Bayesian Network. *Expert Systems with Applications*, *221*, 119749. https://doi.org/10.1016/j.eswa.2023.119749

Harikrishnakumar, R., Nannapaneni, S., Nguyen, N. H., Steck, J. E., & Behrman, E. C. (2020). *A Quantum Annealing Approach for Dynamic Multi-Depot Capacitated Vehicle Routing Problem* (No. arXiv:2005.12478). arXiv. https://doi.org/10.48550/arXiv.2005.12478

Heo, J., Won, K., Yang, H.-J., Hong, J.-P., & Choi, S.-G. (2019). Photonic scheme of discrete quantum Fourier transform for quantum algorithms via quantum dots. *Scientific Reports*, *9*(1), 12440.

Hussain, H., Javaid, M. B., Khan, F. S., Dalal, A., & Khalique, A. (2020). Optimal control of traffic signals using quantum annealing. *Quantum Information Processing*, *19*, 1–18.

IBM. (2023). *IBM Quantum System Two: The era of quantum utility is here | IBM Quantum Computing Blog*. https://www.ibm.com/quantum/blog/quantum-roadmap-2033

Jain, S. (2021). Solving the Traveling Salesman Problem on the D-Wave Quantum Computer. *Frontiers in Physics*, *9*, 760783. https://doi.org/10.3389/fphy.2021.760783

Kadowaki, T. (2002). *Study of Optimization Problems by Quantum Annealing* (No. arXiv:quant-ph/0205020). arXiv. https://doi.org/10.48550/arXiv.quant-ph/0205020

Kadowaki, T., & Nishimori, H. (1998). Quantum annealing in the transverse Ising model. *Physical Review E*, *58*(5), 5355–5363. https://doi.org/10.1103/PhysRevE.58.5355

Kwon, S., Tomonaga, A., Lakshmi Bhai, G., Devitt, S. J., & Tsai, J.-S. (2021). Gate-based superconducting quantum computing. *Journal of Applied Physics*, *129*(4), 041102.

Leymann, F., & Barzen, J. (2020). The bitter truth about gate-based quantum algorithms in the NISQ era. *Quantum Science and Technology*, *5*(4), 044007.

Li, H.-S., Fan, P., Xia, H., Song, S., & He, X. (2018). The quantum Fourier transform based on quantum vision representation. *Quantum Information Processing*, *17*(12), 333.

Li, H.-S., Song, S., Fan, P., Peng, H., Xia, H., & Liang, Y. (2019). Quantum vision representations and multi-dimensional quantum transforms. *Information Sciences*, *502*, 42–58.

Majumder, R., Khan, S. M., Ahmed, F., Khan, Z., Ngeni, F., Comert, G., Mwakalonge, J., Michalaka, D., & Chowdhury, M. (2021). *Hybrid Classical-Quantum Deep Learning Models for Autonomous Vehicle Traffic Image Classification Under Adversarial Attack* (No. arXiv:2108.01125). arXiv. https://doi.org/10.48550/arXiv.2108.01125

Martín-Guerrero, J. D., & Lamata, L. (2022). Quantum Machine Learning: A tutorial. *Neurocomputing*, *470*, 457–461. https://doi.org/10.1016/j.neucom.2021.02.102

Mohanty, N., Behera, B. K., & Ferrie, C. (2023). Analysis of the Vehicle Routing Problem Solved via Hybrid Quantum Algorithms in the Presence of Noisy Channels. *IEEE Transactions on Quantum Engineering*, *4*, 1–14. https://doi.org/10.1109/TQE.2023.3303989

Neukart, F., Compostella, G., Seidel, C., Von Dollen, D., Yarkoni, S., & Parney, B. (2017). Traffic flow optimization using a quantum annealer. *Frontiers in ICT*, *4*, 29.

Nielsen, M. A., & Chuang, I. L. (2010). *Quantum computation and quantum information*. Cambridge university press.

Niu, C., Rastogi, P., Soman, J., Tamuli, K., & Dixit, V. V. (2024). *Applying Quantum Computing to Solve Multicommodity Network Flow Problem* (No. arXiv:2402.04758). arXiv. https://doi.org/10.48550/arXiv.2402.04758

Osaba, E., Villar-Rodriguez, E., Oregi, I., & Moreno-Fernandez-de-Leceta, A. (2021). Hybrid Quantum Computing - Tabu Search Algorithm for Partitioning Problems: Preliminary Study on the Traveling Salesman Problem. *2021 IEEE Congress on Evolutionary Computation (CEC)*, 351–358. https://doi.org/10.1109/CEC45853.2021.9504923

Papalitsas, C., Andronikos, T., Giannakis, K., Theocharopoulou, G., & Fanarioti, S. (2019). A QUBO model for the traveling salesman problem with time windows. *Algorithms*, *12*(11), 224.

Park, S., Kim, J. P., Park, C., Jung, S., & Kim, J. (2024). Quantum Multi-Agent Reinforcement Learning for Autonomous Mobility Cooperation. *IEEE Communications Magazine*, 1–7. https://doi.org/10.1109/MCOM.020.2300199

Pearson, A., Mishra, A., Hen, I., & Lidar, D. A. (2019). Analog errors in quantum annealing: Doom and hope. *Npj Quantum Information*, *5*(1), 107. https://doi.org/10.1038/s41534-019-0210-7

Qu, Z., Liu, X., & Zheng, M. (2023). Temporal-Spatial Quantum Graph Convolutional Neural Network Based on Schrödinger Approach for Traffic Congestion Prediction. *IEEE Transactions on Intelligent Transportation Systems*, *24*(8), 8677–8686. https://doi.org/10.1109/TITS.2022.3203791

Sales, J. F. A., & Araos, R. A. P. (2023). *Adiabatic Quantum Computing for*

*Logistic Transport Optimization* (No. arXiv:2301.07691). arXiv. https://doi.org/10.48550/arXiv.2301.07691

Schuld, M., Sinayskiy, I., & Petruccione, F. (2015). An introduction to quantum machine learning. *Contemporary Physics*, *56*(2), 172–185. https://doi.org/10.1080/00107514.2014.964942

Shor, P. W. (1994). Algorithms for quantum computation: Discrete logarithms and factoring. *Proceedings 35th Annual Symposium on Foundations of Computer Science*, 124–134.

Sinaga, T. J. H., Anwar, K., Amalia, N., Sunnardianto, G. K., & Budiman, G. (2023). Important Quantum Gates for Quantum Algorithms of Travelling Salesman Problem. *2023 International Conference on Artificial Intelligence, Blockchain, Cloud Computing, and Data Analytics (ICoABCD)*, 146–151. https://doi.org/10.1109/ICoABCD59879.2023.10390921

Stollenwerk, T., O'Gorman, B., Venturelli, D., Mandrà, S., Rodionova, O., Ng, H. K., Sridhar, B., Rieffel, E. G., & Biswas, R. (2020). Quantum Annealing Applied to De-Conflicting Optimal Trajectories for Air Traffic Management. *IEEE Transactions on Intelligent Transportation Systems*, *21*(1), 285–297. https://doi.org/10.1109/TITS.2019.2891235

Suen, W. Y., Yat Lee, C., & Lau, H. C. (2021). Quantum-inspired algorithm for Vehicle Sharing Problem. *2021 IEEE International Conference on Quantum Computing and Engineering (QCE)*, 17–23. https://doi.org/10.1109/QCE52317.2021.00017

Xu, H.-Z., Chen, J.-H., Zhang, X.-C., Lu, T.-E., Gao, T.-Z., Wen, K., & Ma, Y. (2023). High-speed train timetable optimization based on space–time network model and quantum simulator. *Quantum Information Processing*, *22*(11), 418. https://doi.org/10.1007/s11128-023-04170-3

Yamany, W., Moustafa, N., & Turnbull, B. (2023). OQFL: An Optimized Quantum-Based Federated Learning Framework for Defending Against Adversarial Attacks in Intelligent Transportation Systems. *IEEE Transactions on Intelligent Transportation Systems*, *24*(1), 893–903. https://doi.org/10.1109/TITS.2021.3130906